# Encounters of star clusters as a possible source of disturbances in their internal structure

**M. D. Sizova[1], L. A. Maksimova[2], S. V. Vereshchagin[1,*]**

[1]Institute of Astronomy, Russian Academy of Sciences, Moscow, 119017 Russia
[2]Research Laboratory for Astrochemistry, Ural Federal University, Kuibysheva St. 48, Yekaterinburg 620026, Russia
*e-mail: svvs@ya.ru

**Abstract.** We searched for past close encounters of the Hyades and Pleiades with all 5736 clusters of the Hunt & Reffert catalog, integrating orbits with galpy over the last 100 Myr and validating candidates by Monte Carlo. Most nominally close pairs are unstable under catalog errors; only two remain reliable: Pleiades and HSC 751 and Hyades and HSC 2986. HSC 2986 coincides with the star-forming cloud Corona Australis, which the Hyades crossed about 5 Myr ago, passing ~11 pc from its dense core. Such interactions may be the cause of the mini-clusters discovered in the Pleiades.

## 1. INTRODUCTION

The approach of star clusters is a short-time gravitational interaction that can change the structure of the participants. These events play an important role in the evolution of clusters, the formation of binary and multiple clusters. We can identify the physical mechanisms that will operate in this case, partially discussed in the works of Zhu, Li, Zhang (2026) [2], Loibnegger B. et al. (2021) [3]. Tidal disruption and "evaporation" are the main mechanisms. Even if the clusters do not collide directly, the gravitational field can cause powerful tidal forces. They can deform and "stretch" the outer parts of the open cluster, as well as destabilize the orbits of stars within the clusters. On the other hand, the process of approach itself can accelerate the "evaporation" of the stars from the cluster. Not only can a cluster change its orbit within the Galaxy, but kinematic groups can also be released in the outer regions of the cluster (for example, in the corona), which can form a moving group within the cluster.

In our era, paired clusters are not uncommon; a catalog of such objects has been published, for example, in the paper by Palma et al. (2025) [4]. The catalog of these authors (VizieR: J/A+A/693/A218) contains a complete data set, including all 617 paired and 261 group systems. Grinenko and Kovaleva (2025) showed that mutual approaches of open clusters at distances on the order of their sizes are not uncommon: in the solar neighborhood, their frequency is 35–40 events per 1 million years. This work estimates the frequency of approaches for the cluster ensemble as a whole; the history of close passages of the Hyades and Pleiades specifically and the relationship of their modern structure with individual partner clusters have not been systematically studied.

Maxwell (2025) [5] discovered stellar groups in the Pleiades OSC. The network analysis Gaia DR3 astrometry of the Pleiades help to reveal the cluster’s internal kinematic structure. Using a Mahalanobis distance graph in proper-motion space was identified the largest connected component of 1556 stars at a threshold of $\tau = 4.0$ and partitions this network into eight subgroups of sizes 741, 629, 75, 55, 39, 6, 6, and 5. This method to provides a compact way to characterize substructure in nearby open clusters using Gaia astrometry.

In this paper, we consider the close encounters of the Hyades and Pleiades with clusters in the Hunt & Reffert (2021) catalog [1], select pairs that are stable given observational errors, and discuss the most notable finding—the close encounter of the Hyades near the star-forming region Corona Australis. Section 2 describes the data used. Section 3 presents the orbital calculation method and Monte Carlo validation. Section 4 presents the results of the screening and selection of reliable pairs. Section 5 provides a discussion and conclusions.

## 2. DATA

The Hunt & Reffert (2021) catalog of star clusters (URL=https://ui.adsabs.harvard.edu/abs/2023A%26A...673A.114H), constructed using Gaia data, was used. After standard filters (excluding objects without radial velocities and obviously erroneous entries), a list of n = 5736 clusters was included in the calculations. Approaches of the Hyades and Pleiades were searched for with all these clusters; a preliminary selection based on the present-day distance from the Sun $d_0$ was not performed, since a cluster that is now distant could have passed close in the past. For each cluster, the equatorial coordinates, parallax, proper motions, and radial velocity with their catalog errors were used, as well as estimates of the age, mass, and radius r50 (the radius containing half the mass). Note that the Theia 3475 is excluded from consideration, since the radial velocity for it is RV = 664 km/s.

## 3. METHOD

The motion of clusters and the Sun was integrated into the gravitational field of the Galaxy using the *galpy* package by Bovy (2015). The MWPotential2014 potential was adopted, including the bulge, disk (according to Miyamoto–Nagai) and halo (according to the Navarro–Frank–White profile), the distance of the Sun from the Galactic center $R_0$ = 8.178 kpc, the velocity of the local standard of rest $V_0$ = 233.1 km/s, the peculiar velocity of the Sun is (11.1, 245.04, 7.25) km/s and its height above the disk plane $Z_0$ = 20.8 pc.

The orbits were numerically integrated backward in time from the present epoch over the interval from 0 to −100 Myr with a step of dt = 0.1 Myr (1000 steps). For each partner cluster, the time interval was additionally constrained by its age, i.e., close encounters earlier than the cluster's formation were not considered. At each step, the distance between the Hyades (Pleiades) and each of the catalog clusters was calculated; the minimum of these distances was taken as $d_{min}$, and the corresponding instant was taken as $t_{min}$. In the first step, clusters were treated as point objects (ignoring tidal radii). Pairs with $d_{min}$ < 150 pc were retained for analysis, and pairs with $d_{min}$ < 30 pc were selected as close encounter candidates for detailed examination.

To calculate the reliability of the found approach parameters, Monte Carlo simulations were performed. For each candidate pair, 2000 orbital realizations were constructed, in which the parallax, proper motions, and radial velocity were varied within the catalog errors. The mean $d_{min}$ and its standard deviation (±σ) were determined from the ensemble of realizations. We consider a pair reliable if its mean $d_{min}$ according to Monte Carlo remains close, less than 25 pc (pairs with a mean of 25–35 pc are treated as borderline, and those above 35 pc are rejected). That is, the close approach itself is not destroyed when accounting for observational errors. If, when varying the input data, the mean $d_{min}$ increases to hundreds of parsecs, the approach is considered an artifact of the nominal orbit. We emphasize that we used the preservation of a small $d_{min}$ , rather than a narrow confidence interval. A wide interval with a small mean still corresponds to a real close passage.

## 4. RESULTS

At the stage of searching by orbital elements for both clusters, a number of close passes with $d_{min}$ < 45 pc were found (Fig. 1). The closest candidates turned out to be for the Pleiades – HSC 751 ($d_{min}$ = 8.3 pc, $t_{min}$ = −11.0 Myr) and for the Hyades – HSC 2986 ($d_{min}$ = 11.0 pc, $t_{min}$ = −5.1 Myr). The complete lists of candidates for both clusters are given in Table 1 and Table 2, respectively. Note that there are no intersections among the candidates ($d_{min}$ < 45 pc): the Hyades and Pleiades approached different clusters, and they did not have common partners at these distances.

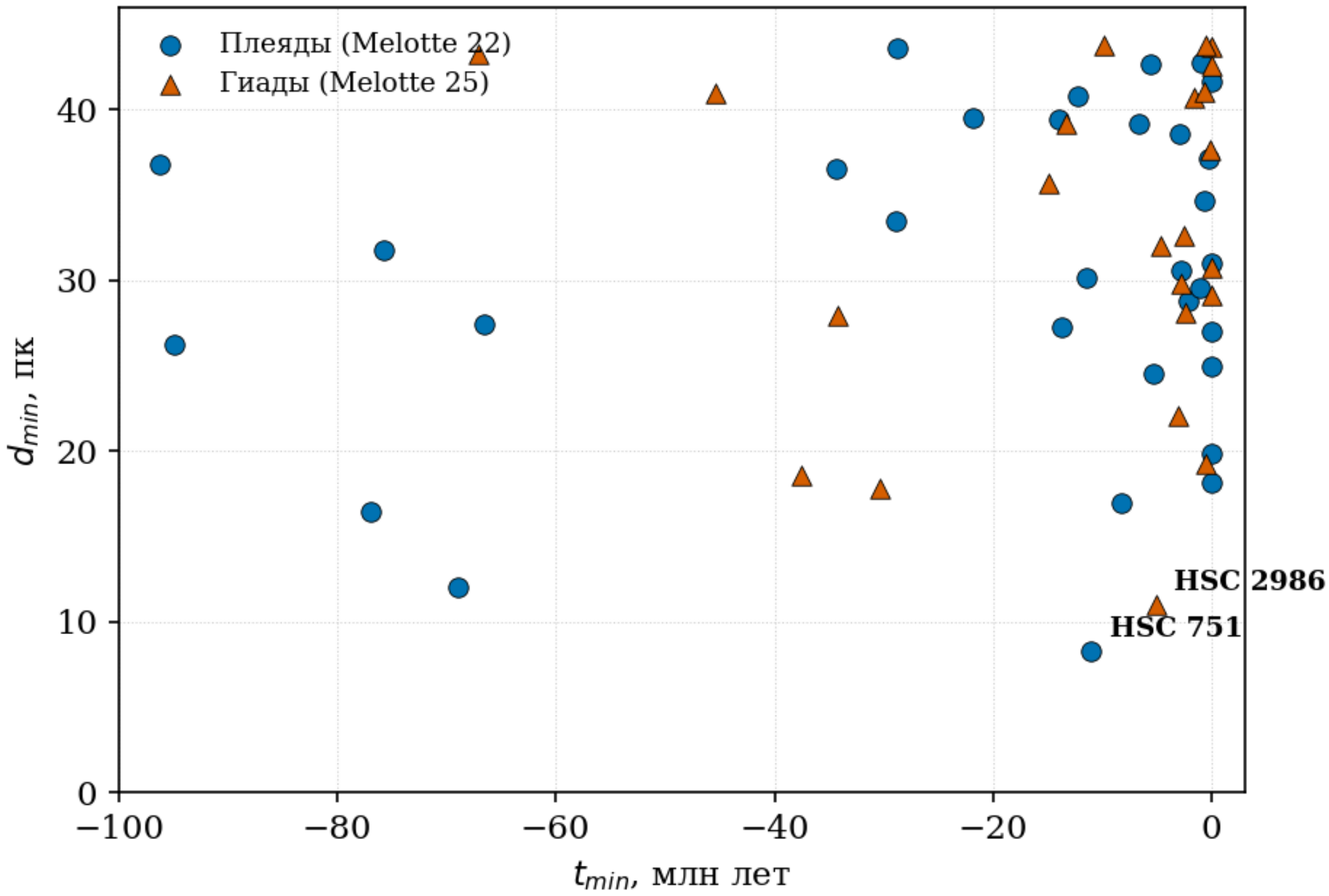

Fig. 1. Minimum distances $d_{min}$, and corresponding moments $t_{min}$ for cluster candidates for encounters with the Pleiades (blue circles) and Hyades (orange triangles) along calculated orbits. The closest candidates for each cluster are labeled.

Table 1. Parameters of approach to the Pleiades (cluster 1) of clusters whose names are in the column “clusters 2” and $d_{min}$<30 pc.

| Star cluster 2 | $d_{min}$, pc | $t_{min}$, Myr | $d_0$, pc | age, Myr |
|---|---|---|---|---|
| HSC 751 | 8.3 | -11.0 | 327 | 184 |
| Collinder 394 | 12.0 | -68.9 | 808 | 84 |
| Theia 89 | 16.4 | -77.0 | 839 | 686 |
| Teutsch J0450.9+5209 | 17.0 | -8.2 | 194 | 283 |
| HSC 1340 | 18.1 | 0.0 | 18 | 27 |
| HSC 1318 | 19.8 | 0.0 | 20 | 6 |
| HSC 1254 | 24.5 | -5.4 | 54 | 115 |
| Theia 7 | 25.0 | 0.0 | 25 | 187 |
| CWNU 2682 | 26.2 | -94.9 | 4207 | 270 |
| HSC 1373 | 27.0 | 0.0 | 27 | 44 |
| Zanin 6 | 27.3 | -13.7 | 328 | 125 |
| UPK 303 | 27.4 | -66.6 | 101 | 111 |
| HSC 911 | 28.8 | -2.2 | 101 | 4377 |
| HSC 1195 | 29.5 | -1.1 | 45 | 253 |

| Star cluster 2 | $d_{min}$, pc | $t_{min}$, Myr | $d_0$, pc | age, Myr |
|---|---|---|---|---|
| HSC 1040 | 30.1 | -11.4 | 207 | 76 |
| HSC 1368 | 30.6 | -2.8 | 64 | 47 |
| Theia 66 | 31.0 | 0.0 | 31 | 6 |
| NGC 6204 | 31.7 | -75.7 | 1278 | 80 |
| OCSN 36 | 33.5 | -28.9 | 414 | 229 |
| HSC 1288 | 34.6 | -0.7 | 38 | 147 |
| UBC 136 | 36.5 | -34.3 | 1396 | 105 |
| Teutsch 28 | 36.8 | -96.3 | 3267 | 211 |
| CWNU 1129 | 37.2 | -0.3 | 38 | 15 |
| HSC 1250 | 38.5 | -3.0 | 158 | 4 |
| HSC 976 | 39.1 | -6.7 | 148 | 14 |
| HSC 1054 | 39.4 | -14.0 | 274 | 155 |
| ASCC 123 | 39.5 | -21.9 | 409 | 1064 |
| HSC 745 | 40.8 | -12.2 | 373 | 116 |
| HSC 1438 | 41.6 | 0.0 | 42 | 57 |
| Melotte 20 | 42.7 | -5.6 | 78 | 56 |
| Theia 54 | 42.8 | -1.0 | 47 | 4 |
| HSC 1651 | 43.6 | -28.7 | 3221 | 200 |

Table 1. Parameters of approach to the Hyades (cluster 1) of clusters whose names are in the column “clusters 2” and $d_{min}$<45 pc.

| Star cluster 2 | $d_{min}$ , pc | $t_{min}$, Myr | $d_0$, pc | age, Myr |
|---|---|---|---|---|
| HSC 2986 | 11.0 | -5.1 | 187 | 7 |
| HSC 2743 | 17.8 | -30.3 | 1220 | 569 |
| HSC 207 | 18.6 | -37.5 | 488 | 58 |
| FSR 1017 | 19.3 | -0.6 | 23 | 16 |
| HSC 396 | 22.0 | -3.1 | 123 | 135 |
| HSC 2411 | 27.9 | -34.2 | 1382 | 208 |
| HSC 2846 | 28.1 | -2.4 | 88 | 54 |
| HSC 1470 | 29.2 | 0.0 | 29 | 447 |
| Theia 3475 | 29.8 | -2.8 | 1761 | 145 |
| HSC 1566 | 30.7 | 0.0 | 31 | 115 |
| HSC 2900 | 32.1 | -4.7 | 185 | 7 |
| HSC 103 | 32.6 | -2.6 | 145 | 4 |
| Theia 240 | 35.6 | -14.9 | 452 | 110 |
| HSC 1152 | 37.7 | -0.1 | 38 | 131 |
| Collinder 350 | 39.2 | -13.3 | 408 | 247 |
| beta Tuc Group | 40.7 | -1.6 | 72 | 31 |
| HSC 2886 | 41.0 | -45.4 | 1588 | 141 |
| HSC 1900 | 41.0 | -0.7 | 47 | 27 |
| HSC 1347 | 42.6 | 0.0 | 43 | 246 |
| Ruprecht 91 | 43.3 | -67.1 | 1046 | 149 |
| HSC 1580 | 43.7 | 0.0 | 44 | 132 |
| HSC 180 | 43.7 | -9.9 | 260 | 285 |
| HSC 906 | 43.8 | -0.6 | 57 | 277 |

Monte Carlo simulations showed that most of the close pairs found are unstable (Fig. 2). On timescales of ~70 million years, a spatial velocity error of about 1 km/s leads to a position deviation of ~100 pc, so pairs with large $t_{min}$ and large initial distances $d_0$, "move apart" when the input data are varied. For example, the Pleiades – Collinder 394 pair, which is close in nominal orbit ($d_{min}$ = 12 pc), yields a mean $d_{min}$ = 235 pc when accounting for errors.

Only two pairs turned out to be reliable. For the Pleiades, this is HSC 751: mean $d_{min}$ = 11.8 ± 8.9 pc at $t_{min}$ = −11.0 Myr. For the Hyades, this is HSC 2986: mean $d_{min}$ = 10.2 ± 0.6 pc at $t_{min}$ = −5.1 Myr.

It is important to distinguish the reliability criterion from the width of the error interval: reliability is set by how small and robust the closest approach is, not by how narrow the Monte Carlo scatter is. For the Pleiades–HSC 751 pair the scatter is wide, but its $d_{min}$ stays well below 25 pc (mean 11.8 ± 8.9 pc), so we classify it as a reliable encounter; for the Pleiades–Zanin 6 pair the scatter is narrow but sits right at the screening threshold (mean 28.3 ± 3.3 pc), so it is treated as borderline. In Fig. 2 three zones are shown — reliable ($d_{min}$ < 25 pc), borderline (25–35 pc) and rejected (> 35 pc); HSC 751 and HSC 2986 fall in the reliable zone, while Zanin 6 and UPK 303 lie in the borderline band.

Table 3. Monte Carlo results (2000 realizations) for the tested pairs: mean $d_{min}$, ±σ boundaries, mean $t_{min}$, , and reliability conclusion. The pairs Pleiades – HSC 751 and Hyades – HSC 2986 are reliable.

| cluster 1 | cluster 2 | $< d_{min} >$, пк | −σ | +σ | $< t_{min} >$, Myr | comment |
|---|---|---|---|---|---|---|
| Pleiades | HSC 751 | 11.8 | 2.9 | 20.7 | -11.0 | reliable |
| Pleiades | Collinder 394 | 235.5 | 81.6 | 389.4 | -66.8 | rejected M-C |
| Pleiades | Theia 89 | 99.3 | 40.6 | 158.0 | -76.5 | rejected |
| Pleiades | Zanin 6 | 28.3 | 25.0 | 31.6 | -13.8 | borderline |
| Pleiades | UPK 303 | 33.1 | 12.9 | 53.3 | -58.1 | borderline |
| Hyades | HSC 2986 | 10.2 | 9.6 | 10.8 | -5.1 | reliable |
| Hyades | HSC 2743 | 150.0 | 95.2 | 204.8 | -31.5 | rejected |
| Hyades | HSC 207 | 225.3 | 30.4 | 420.2 | -18.3 | rejected |
| Hyades | HSC 2411 | 244.8 | 0.0 | 515.5 | -35.3 | rejected |

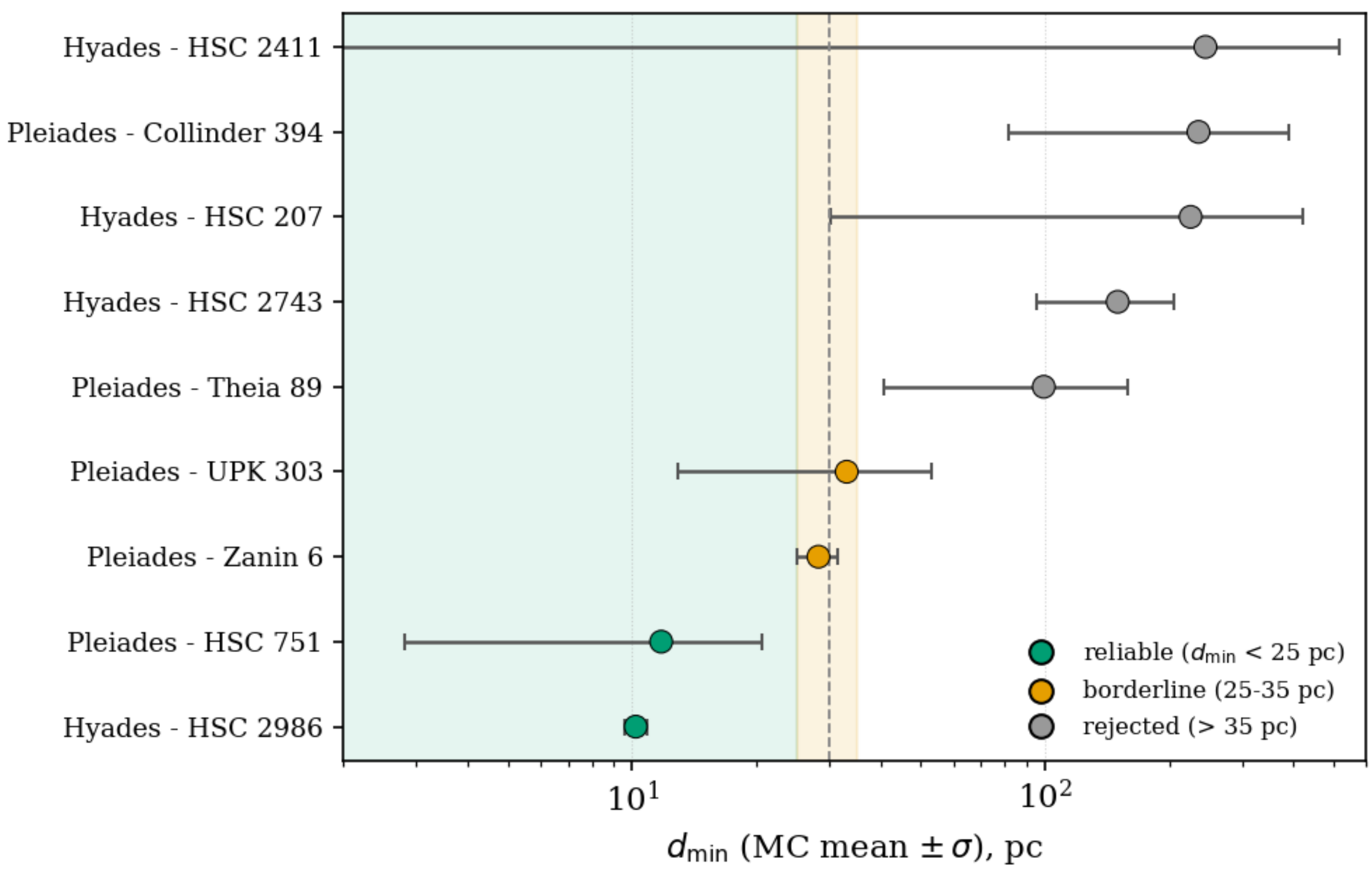

Fig. 2. Mean values of $d_{min}$ with their ±σ Monte Carlo scatter for different pairs (logarithmic scale). Reliable pairs (< 25 pc) are highlighted in green, borderline pairs (25–35 pc) in orange, and rejected pairs (> 35 pc) in gray. The dashed line is the nominal 30 pc screening threshold.

The most remarkable finding is for the Hyades. The object HSC 2986 has galactic coordinates l = 360.0°, b = −17.7° at a distance of 148 pc from the Sun—this exactly coincides with the position of the molecular cloud Corona Australis, the nearest active star-forming region. The peculiarity of HSC 2986 is that it is a young object with an age of only 7.1 Myr. According to our integration of orbits, at the time of its formation, HSC 2986 was located ≈75 pc from the Hyades, and about 2 Myr later, about 5.1 Myr ago, the distance between them decreased to a minimum of ≈10 pc (Fig. 3). Thus, at the time of close encounter, the age of HSC 2986 was only about 2 million years.

Within 80 pc of Corona Australis at the time of the close encounter were several young clusters (HSC 2900, HSC 103, Theia 67, and HSC 67) with a median age of 7.3 Myr, consistent with the picture of a single complex. The approach of the Hyades to this region and the relative positions of the objects at the time of the close encounter are shown in Fig. 3.

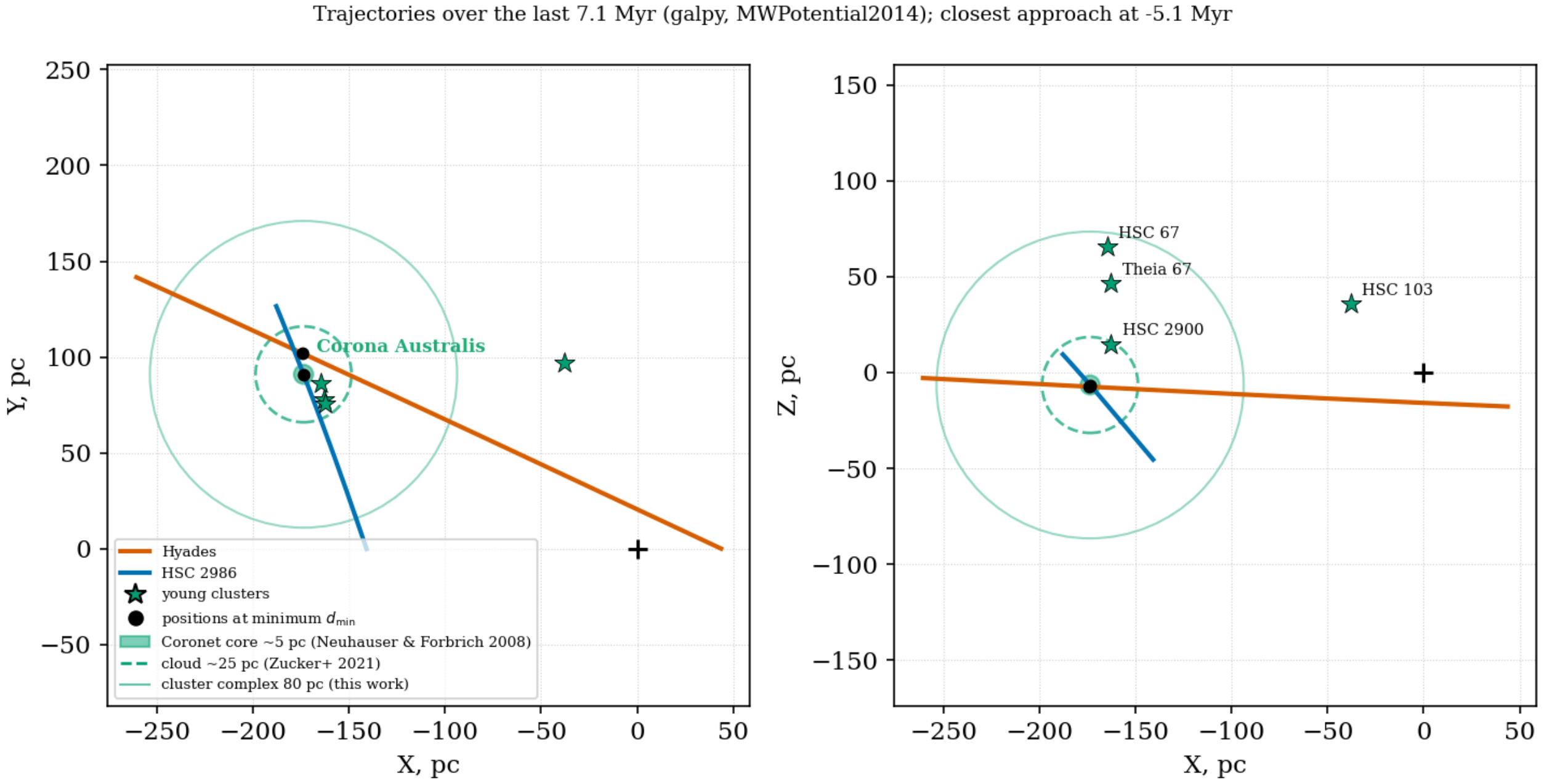

Fig. 3. Trajectory of the Hyades (orange line) and HSC 2986 (blue line) in the heliocentric galactic coordinate system over the past 7.1 Myr. The left panel is the X–Y projection, the right panel is the X–Z projection. Filled circles are the present-day positions, black dots are the positions of the Hyades and HSC 2986 at the time of closest approach (−5.1 Myr). Corona Australis is shown in green at three scales with the source of the sizes indicated: filled circle ≈5 pc is the dense star-forming core of (Neuhauser & Forbrich 2008 [7]), dashed circle ≈25 pc is the molecular cloud (Zucker et al. 2021 [8]), thin outline 80 pc is the vicinity of young clusters of the complex (our estimate). The Hyades track passes through the cloud (≈25 pc), but bypasses the dense core (≈11 pc away). The stars indicate the complex's young clusters (their names are labeled in the right panel), and the cross indicates the Sun.

The change in the distance between the Hyades and HSC 2986 over the entire interval of the latter's existence is shown in Fig. 4: at the present epoch the clusters are separated by ≈187 pc, by the time of the close approach about 5.1 Myr ago the distance is reduced to a minimum of ≈11 pc, and at the time of the formation of HSC 2986 (≈7.1 Myr ago) it was ≈75 pc.

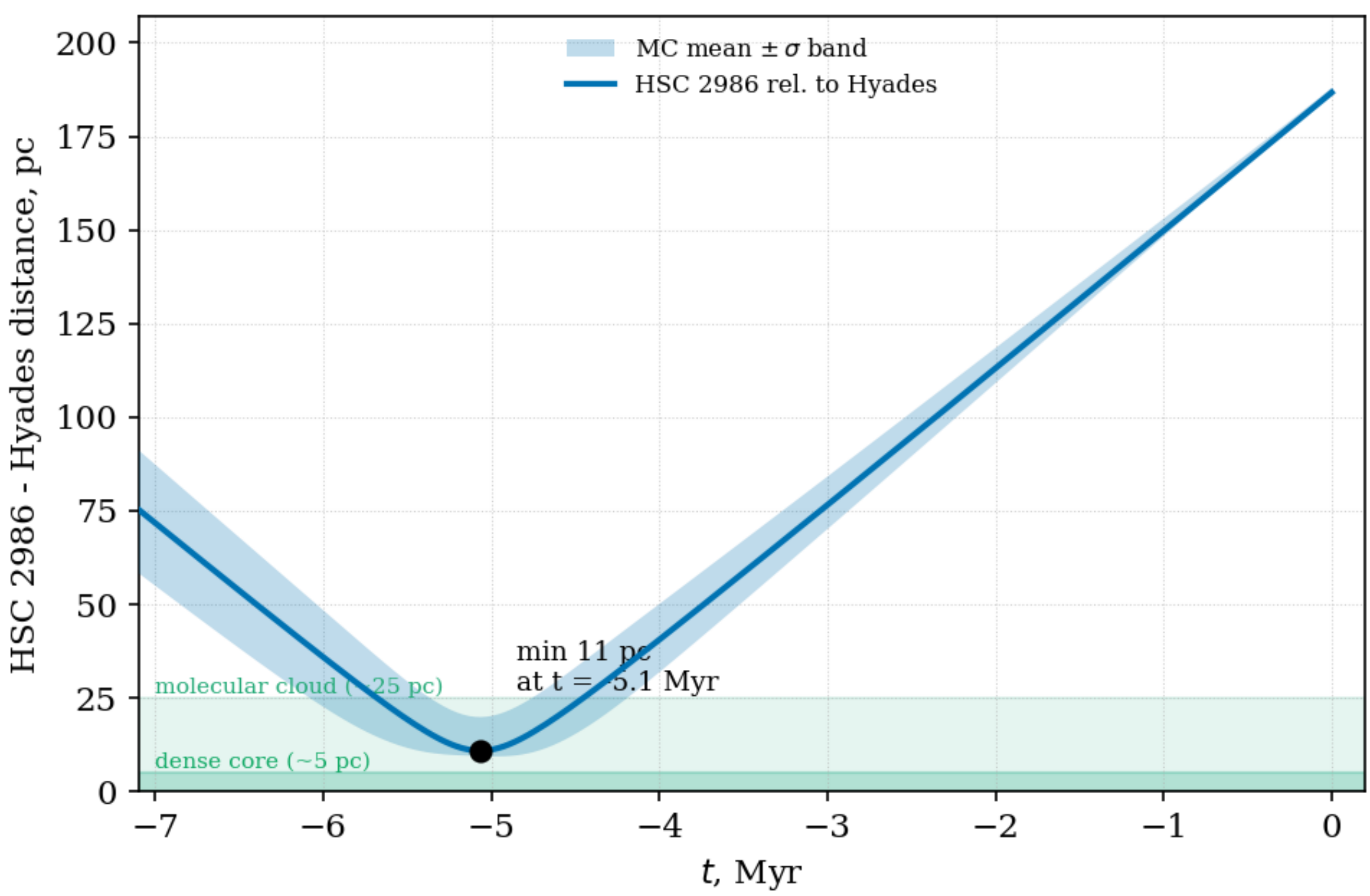

Fig. 4. The distance between HSC 2986 and the Hyades as a function of time over the lifetime of HSC 2986 (0 to −7.1 Myr); the motion of HSC 2986 relative to the Hyades is shown. The shaded band is the ±σ Monte Carlo scatter (sampling HSC 2986 within its catalog errors); the horizontal stripes mark the scales of the molecular cloud (~25 pc) and the dense core (~5 pc) of Corona Australis. The closest approach is ≈11 pc at approximately −5.1 Myr.

The abrupt nature of the encounter (Fig. 4) is explained by the pair's high relative velocity. The Hyades are moving at ≈46 km/s, while HSC 2986 is moving at ≈20 km/s, and their relative velocity is ≈36 km/s. At this velocity, the distance between the objects changes by approximately 36 pc per million years, so the close encounter is short, producing a narrow minimum in d(t).

HSC 2986 движется вместе со своим родительским комплексом. Его пространственная скорость (≈20 км/с) и ее направление близки к скоростям соседних молодых скоплений Corona Australis (HSC 2900, Theia 67 — 16–21 км/с со сходными компонентами V и W). То есть HSC 2986 кинематически близок комплексу, в котором, возможно, родился. Резкое сближение на Рис. 4 обусловлено, возможно, близким прохождением Гиад. Гиады движутся по совершенно иной галактической орбите с большой компонентой скорости U ≈ −42 км/с. Сама область Corona Australis при этом за 5 млн лет сместилась приблизительно на 100 пк. На Рис. 3 показано современное положение области, и положение на момент встречи).

HSC 2986 moves together with its parent complex. Its spatial velocity ≈20 km/s and its direction are close to the velocities of neighboring young Corona Australis clusters (HSC 2900, Theia 67 – 16–21 km/s with similar V and W components). That is, HSC 2986 is kinematically close to the complex in which it was possibly born. The sharp approach in Fig. 4 is possibly due to the close passage of the Hyades. The Hyades move along a completely different galactic orbit with a large velocity component U ≈ −42 km/s. The Corona Australis region itself has shifted by approximately 100 pc over 5 million years. Figure 3 shows the current position of the region and its position at the time of the encounter.

## 5. DISCUSSION

It is useful to compare our results with the independent work of Grinenko & Kovaleva (2025) [5], where cluster approaches were searched for with the same instrument (galpy, MWPotential2014 potential) using the catalog [1] (≈5749 clusters with known radial velocities). Their calculation was performed with a step of 2 million years over an interval of ±32 million years, and the published catalog is a forecast of future approaches (490 pairs for 32 million years ahead, 29 pairs in closest proximity at the present time); the emphasis is on the statistics of the approach frequency, specific orbital energy and the difference in ages of the pairs, up to the search for binary cluster candidates. Our work, in contrast, is aimed at a detailed history of the Hyades and Pleiades in the past, with a longer interval of 100 million years and a finer step (0.1 million years). Where the regions overlap—at the present epoch—the results agree: for example, both studies give a minimum distance of ≈19.8 pc for the Pleiades–HSC 1318 pair at the present time. The estimate of the frequency of close cluster encounters obtained in [5] (35–40 events per 1 million years in the solar neighborhood) confirms the plausibility of the two past passages we found. However, our main pairs (HSC 751 at −11 million years and HSC 2986 at −5.1 million years) belong to the past and are therefore not included in their predicted catalog of future encounters—in this sense, the two studies complement each other.

Moreover, both of our key findings are missing in [5] for a methodological reason—due to time aliasing on the coarse grid. With an integration step of 2 Myr, the grid nodes for the Pleiades–HSC 751 pair fall at −10 and −12 Myr, where the distance between the clusters is ≈40 and ≈27 pc, while the true minimum (≈8 pc) is reached between the nodes, at about −11 Myr. Similarly, for the Hyades–HSC 2986 pair, the nodes at −4 and −6 Myr yield ≈41 and ≈36 pc, and the minimum of ≈10 pc falls at −5.1 Myr, exactly between the nodes. In both cases, both neighboring nodes lie above the catalog inclusion threshold, so the rapid, sharp encounter "falls" through the coarse grid and is not registered. For the same clusters, Grinenko & Kovaleva [5] find only slower transits that fall on the nodes of their grid: for the Pleiades, these are HSC 1318 (t = 0, 19.8 pc) and UPK 535 (t = +12 Myr, 15.6 pc), and for the Hyades, Theia 65 (t = +2 Myr, 6.7 pc). Thus, the small step size dt = 0.1 Myr is an advantage of our approach: it resolves short, close encounters that are invisible with a step size of 2 Myr, and this is why the pairs HSC 751 and HSC 2986 were detected for the first time.

The sizes of the approaching clusters deserve special consideration, as they determine whether the passage was due to tidal perturbation or physical interpenetration. According to the Hunt & Reffert (2021) catalog, the half-mass radius of r50 is 4.2 pc for the Hyades and 2.9 pc for HSC 2986; for the Pleiades–HSC 751 pair, these are 1.1 and 1.4 pc, respectively. The minimum approach distances (≈10 pc for the Hyades and ≈12 pc for the Pleiades) exceed the sum of the half-mass radii of the partners by several times, meaning the clusters likely did not interpenetrate, but passed at a distance of approximately several times their radii. At such distances, tidal perturbation of the outer parts and tidal tails of the less massive partner may be noticeable. The mass of the Hyades (≈409 solar masses) significantly exceeds the mass of the young HSC 2986 (≈135 solar masses), so the Hyades could have had a stronger impact on HSC 2986 than vice versa.

It should be taken into account that the radius r50 is not constant over time. The structure of clusters evolves as they dynamically evolve and dissipate. According to Gaia observations, the radius of the cluster core decreases on average with age, while the total size (including the halo) increases slightly (Tarricq et al. [9]). The dissipation of clusters in the solar neighborhood was modeled in the paper by Almeida et al. (2025) [10]. For the Hyades, however, the interval of ≈5 million years since the encounter is less than one percent of their age (about 600 million years), so the change in r50 over this time is small and lies within the catalog uncertainties; the use of the modern value of r50 here is justified. An accurate reconstruction of the size of the Hyades at the epoch of the encounter would require N-body modeling and is beyond the scope of this paper.

## 6. CONCLUSIONS

We analyzed the encounters of the Hyades and Pleiades with all clusters in the Hunt & Reffert (2021) [1] catalog using the following approach: numerical calculation and orbital comparison, selection of candidates with $d_{min}$ < 30 pc, and Monte Carlo validation of the selected pairs. The calculations were performed within the adopted simplifications: clusters were treated as point objects, and the integration was limited to a 100-Myr window.

Taking into account the catalog errors, only two pairs remain stable: the Pleiades – HSC 751 and the Hyades – HSC 2986, while the other close passes are unstable over long time intervals. The most interesting pair is the Hyades – HSC 2986. Approximately 5 million years ago, the Hyades passed ≈11 pc from HSC 2986, a young cluster immersed in the Corona Australis cloud. This means that the Hyades passed through an extended molecular cloud (≈25 pc in size according to data on the three-dimensional structure of local clouds [8], but passed its dense core (Coronet, ≈2–3 pc [7]) at a distance of ≈11 pc. In other words, they crossed the cloud complex without touching the regions of the most active star formation.

Clearly, we haven't found all the clusters that have encountered and survived in the Galaxy. The catalog [1] only includes clusters located within 5 kpc of the Sun. There are at least an order of magnitude more of them in the Galaxy. This means there could have been significantly more encounters, even over the 100 Myr time interval we considered.

## ACKNOWLEDGMENTS

This study used the SIMBAD database managed by CDS, Strasbourg (Wenger et al., 2000). Data from the European Space Agency (ESA) Gaia mission (see http://www.cosmos.esa.int/gaia) processed by the Gaia Data Processing and Analysis Consortium (DPAC; http://www.cosmos.esa.int/web/gaia/dpac/consortium) were also used. Funding for DPAC was provided by national institutions, in particular by institutions participating in the Gaia Multilateral Agreement. In this work, we also used the Galactic Dynamics Library (*galpy*) created by Bovy (2015). This study used data from the NASA Exoplanet Archive, which is managed by the California Institute of Technology under a contract with the National Aeronautics and Space Administration as part of the Exoplanet Exploration Program.

## FUNDING

The M. D. Sizova and S. V. Vereshchagina work was supported by the Institute of Astronomy of the Russian Academy of Sciences (INASAN). The L. A. Maksimova work was supported by the Ministry of Science and Higher Education of the Russian Federation, project FEUZ-2023-0003. No additional grants to carry out or direct this particular research were obtained.

## CONFLICT OF INTEREST

The authors of this work declare that they have no conflicts of interest.

## REFERENCES

1. E. L. Hunt, S. Reffert, Astron. and Astrophys. **646**, A104 (2021). Catalogue J/A+A/686/A42 (URL=https://vizier.cds.unistra.fr/viz-bin/VizieR-3?-source=J/A%2bA/673/A114/members)

2. Z. Zhu, Z. Li, S. Zhang, Astron. and Astrophys. **708**, A87, 12 pp. (2026).

3. B. Loibnegger, E. Pilat-Lohinger, M. Zimmermann, S. Clees, Conference paper, 15th Europlanet Science Congress, Published in the EPSC2021 proceedings; article identifier "672" (2021).

4. T. Palma, V. Coenda, G. Baume, C. Feinstein, Astron. and Astrophys. **693**, id. A218, 8 pp. (2025).

5. A.D. Grinenko, D.A. Kovaleva, arXiv:2509.00471v2 (2025).

6. J. Bovy, galpy: A python library for galactic dynamics, Astrophys. J. Suppl. S. 216, 29, (2015).

7. R. Neuhäuser and J. Forbrich. "The Corona Australis Star Forming Region." Handbook of Star Forming Regions: The Southern Sky, edited by B. Reipurth, vol. 2, Astronomical Society of the Pacific, pp. 735–756, (2008).

8. C. Zucker, A. Goodman, J. Alves, S. Bialy, E. W. Koch, et al. G. (2021). Astrophys. J. **919**, 1, 35, (2021).

9. Y. Tarricq, C. Soubiran, L. Casamiquela, A. Castro-Ginard, J. Olivares, N. Miret-Roig, P. A. B. Galli, Astron. and Astrophys. **659**, A59, (2022).

10. D. Almeida, A. Moitinho, S. Moreira, Astron. and Astrophys. **693**, A305, (2025).